\documentstyle[preprint,prb,aps,epsfig,floats]{revtex}
\tightenlines
\begin{document}
\begin{titlepage}

\title{Short range ferromagnetism and spin glass state in $\mathrm{Y_{0.7}Ca_{0.3}MnO_{3}}$}

\author{R. Mathieu and P. Nordblad}
\address{Department of Materials Science, Uppsala University, Box 534, SE -751 21 Uppsala, Sweden}

\author{D. N. H. Nam\thanks{and Department of Materials Science, Uppsala University, Box 534, SE - 751 21 Uppsala, Sweden} and N. X. Phuc}
\address{Institute of Materials Science, NCST, Nghiado - Caugiay - Hanoi, Vietnam}

\author{N. V. Khiem}
\address{Department of Science and Technology, Hongduc University, Thanhhoa, Vietnam}

\date{\today}

\maketitle

\begin{abstract}
Dynamic magnetic properties of $\mathrm{Y_{0.7}Ca_{0.3}MnO_{3}}$ are reported. 
The system appears to attain local ferromagnetic
order at $T_{\mathrm{SRF}} \approx 70$ K. Below this temperature the
low field magnetization becomes history dependent, i.e. the zero
field cooled (ZFC) and field cooled (FC) magnetization deviate
from each other and closely logarithmic relaxation appears at our
experimental time scales (0.3-$10^{4}$ sec). The zero field cooled
magnetization has a maximum at $T_{\mathrm{f}}\approx 30$ K,
whereas the field cooled magnetization continues to increase,
although less sharply, also below this temperature. Surprisingly,
the dynamics of the system shows non-equilibrium spin glass (SG)
features not only below the maximum in the ZFC magnetization, but
also in the temperature region between this maximum and
$T_{\mathrm{SRF}}$. The aging and temperature cycling experiments
show only quantitative differences in the dynamic behavior above and
below the maximum in the ZFC-magnetization; similarly, memory effects are 
observed in both temperature regions. We attribute the high temperature 
behavior to the existence of clusters of short range ferromagnetic order 
below $T_{\mathrm{SRF}}$; the configuration evolves into a conventional 
spin glass state at temperatures below $T_{\mathrm{f}}$.\\
 
\end{abstract}

\end{titlepage}

\newpage

\section {Introduction}

Magnetic frustration resulting from the competing coexistence of ferromagnetic 
double-exchange (DE) and antiferromagnetic superexchange interaction, is 
present in colossal magneto-resistance (CMR) materials. Recently, frustration 
related effects have been observed in the CMR ferromagnet 
$\mathrm{Nd_{0.7}Sr_{0.3}MnO_{3}}$ 
\cite{nam}; also, a reentrant spin-glass (RSG) phase has been evidenced in 
$\mathrm{La_{0.96-x}Nd_{x}K_{0.04}MnO_{3}}$ \cite{roland} using relaxation 
measurements. In a recent paper \cite{wang}, SG-like behavior has been 
advocated in $\mathrm{Y_{0.7}Ca_{0.3}MnO_{3}}$ from a cusp at $T\approx 30$ K 
in the $M_{\mathrm{ZFC}}(T)$ curve and a corresponding frequency dependence 
in the ac susceptibility. In the present work, we have performed low-field
magnetic relaxation and associated temperature cycling measurements at
temperatures above and below the $M_{\mathrm{ZFC}}(T)$ maximum.
The relaxation of the low-frequency ac susceptibility is also studied to 
investigate memory effects in the two temperature regions. Below $T_{f}$ 
the system exhibits features of a true SG state. However, long-time relaxation 
and aging effects are still found at higher temperatures, well above $T_{f}$. 
Additional magnetic hysteresis measurements reveal ferromagnetic
short-range correlations below $T_{SRF}$, suggesting the existence of clusters 
of ferromagnetic order. Memory effects are observed in both regions.

\section {Sample and experiments}

\begin{figure}
\centerline{\epsfig{figure=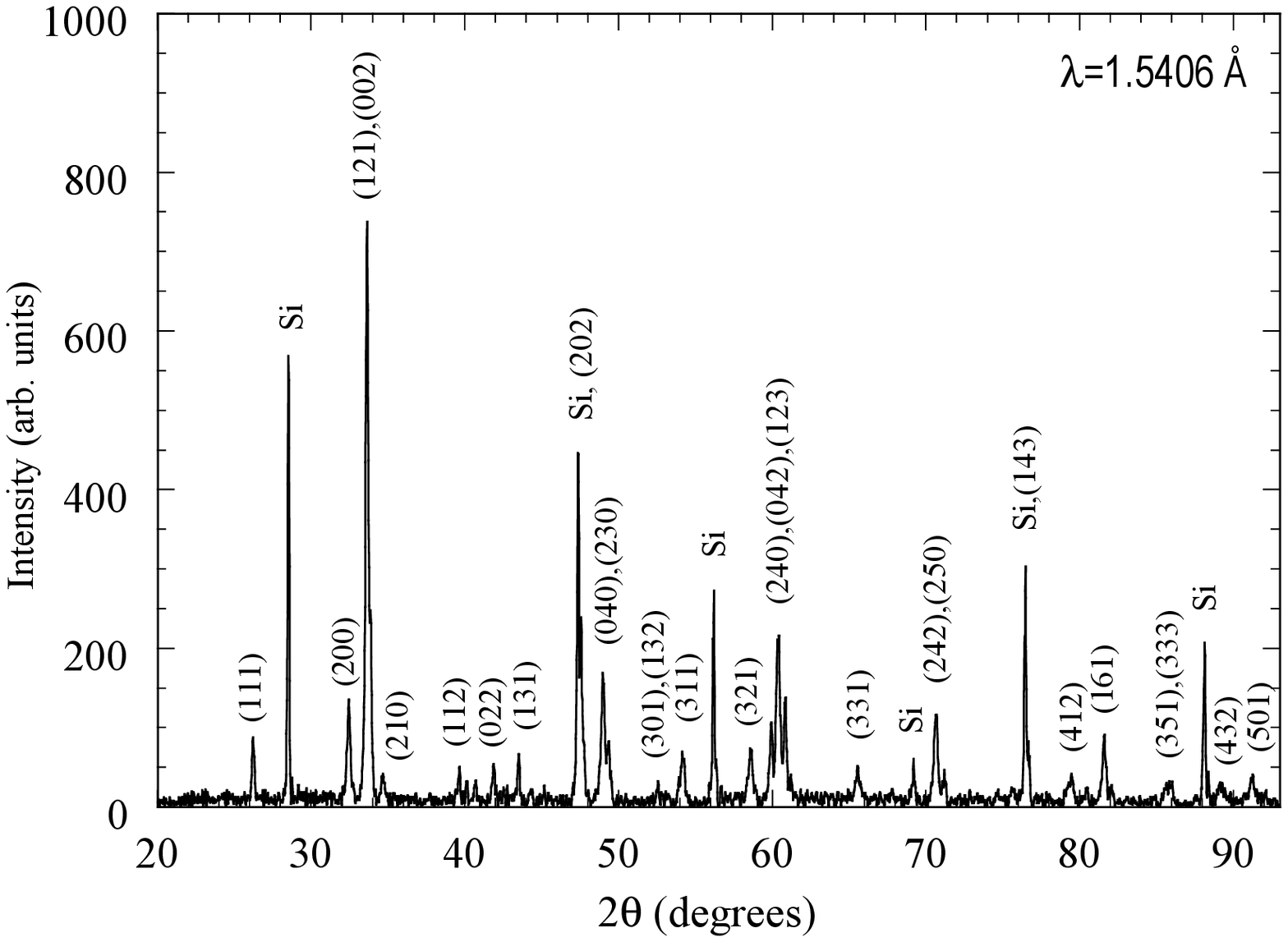,height=6.6cm,width=8cm}}
\caption{Room temperature x-ray diffractogram of the YCMO compound mixed with Si powder. The index of the YCMO reflections is added, and the Si peaks identified.} 
\label{fig0}
\end{figure}

The $\mathrm{Y_{0.7}Ca_{0.3}MnO_{3}}$ (YCMO) compound was prepared by standard solid 
state reaction. After sintering at 1300$^o$ C,
the mixture was annealed in oxygen at 1000, 800, and 600$^o$ C for several days 
at each temperature. The final product was characterized by x-ray diffraction 
technique showing a single phase of orthorhombic structure. The XRD measurements were performed at room temperature using a Siemens D5000 diffractometer with CuK$_{\alpha}$ radiation ($\lambda$=1.5406 \AA) and a scanning step of 0.02$^o$. The sample was first mixed with high-purity Si powder for standard angular calibration. As seen in the diffractogram presented in Fig. \ref{fig0}, which includes the Si peaks, the YCMO reflections can be indexed according to an orthorhombic structure. The obtained lattice parameters are a=5.528 \AA, b=7.441 \AA, and c=5.293 \AA, in agreement with earlier results\cite{wang}. No secondary phases or impurities were detected.\\
The temperature dependence of the zero-field-cooled (ZFC), field-cooled (FC) 
and thermo-remanent (TRM) magnetization, as well as the relaxation of ZFC 
magnetization $m(t)$ and temperature cycling measurements \cite{sandlund} were 
made in a non-commercial low field SQUID system \cite{magnusson}; the 
background field of which is less than 1 mOe. In the relaxation experiments, 
the sample was rapidly cooled in zero field from a reference temperature of 
80 K to a measuring temperature $T_{m}$ and kept there a wait time $t_{w}$. 
After the wait time, a small probing field $H$ was applied and $m(t)$ was
recorded as a function of the time elapsed after the field application. In the 
temperature cycling measurements, just after the wait time and immediately
prior to the application of the probing field, the sample was
additionally subjected to a temperature cycle $\Delta T$ of duration $t_{w2}$. 
Using the same SQUID system, low-frequency ac susceptibility experiments were 
used to investigate  memory phenomena.\\
Additional ``high field'' measurements were performed in a commercial Quantum 
Design MPMS5 SQUID magnetometer (Curie Weiss behavior and Arrot plots) and a 
Lakeshore 7225 susceptometer (ac susceptibility in large superimposed dc-fields).

\section{Results and discussion}

\subsection{Dc measurements}

\begin{figure}
\centerline{\epsfig{figure=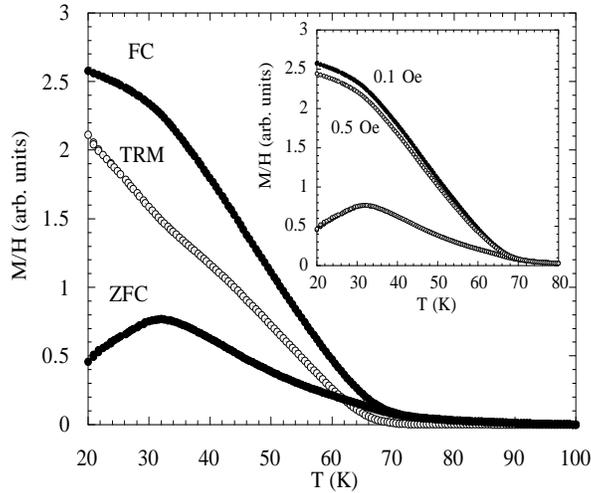,height=6.6truecm,width=8truecm}}
\caption{$M_{ZFC}/H$, $M_{FC}/H$, and $M_{TRM}/H$ as functions of temperature using an applied field of 0.1 Oe. The inset shows $M_{ZFC}/H$ and $M_{FC}/H$ at $H$=0.1 Oe and $H$=0.5 Oe.} 
\label{fig1}
\end{figure}

\begin{figure}
\centerline{\epsfig{figure=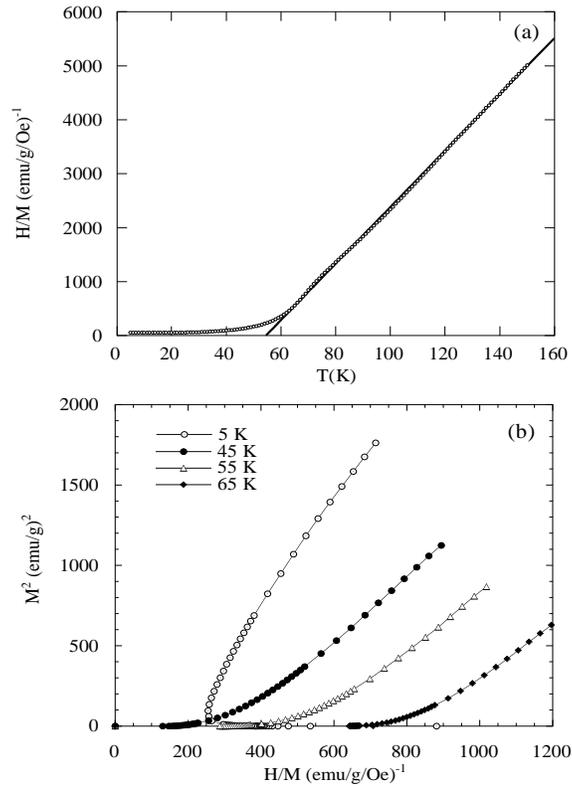,height=10.6cm,width=8cm}}
\caption{(a) Curie-Weiss law from magnetization measurements made in a larger field ($H$=20 Oe) up to higher temperatures. (b) Arrot plots of magnetization curves measured at 5, 45, 55, and 65 K.}
\label{fig2}
\end{figure}

\begin{figure}
\centerline{\epsfig{figure=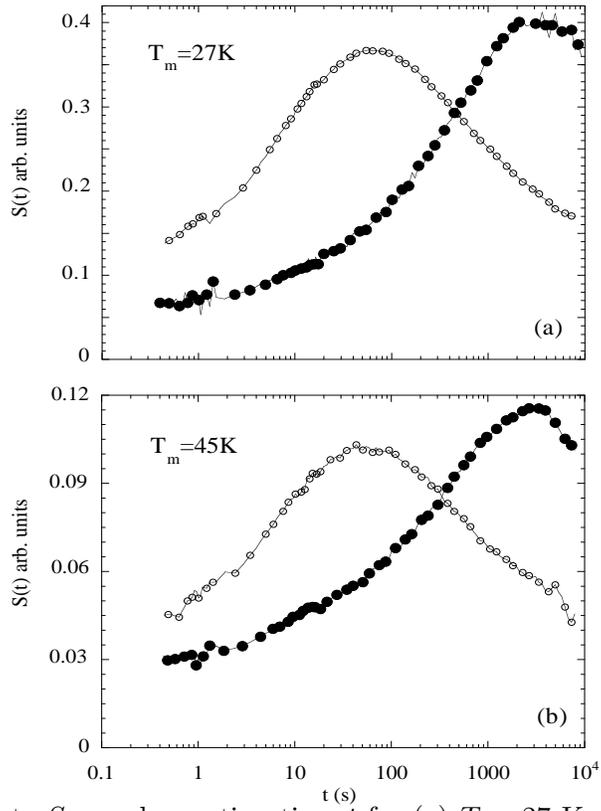,height=10.6cm,width=8cm}}
\caption{Relaxation rate $S$ vs. observation time $t$ for (a) $T=27$ K and (b) $T=45$ K; results for $t_w$= 0s (open symbols) and 3000s (filled symbols) are presented. $H$=0.1 Oe.}
\label{fig3}
\end{figure}

\begin{figure}
\centerline{\epsfig{figure=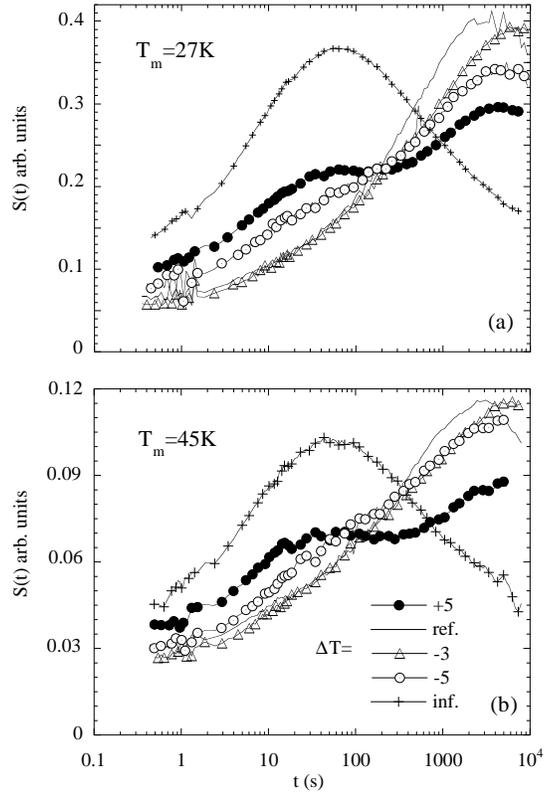,height=10.6cm,width=8cm}}
\caption{Relaxation rate $S$ recorded at (a) $T_{m}=27$ K and (b) 45 K after positive and negative temperature cyclings. $H=0.1$ G and $t_{w1}$=3000s. $t_{w2}$=30000s for negative cycles; $t_{w2}$=0s for the positive ones. Infinite $\Delta T$ corresponds to $t_{w1}$=0s.} 
\label{fig4}
\end{figure}

Figure \ref{fig1} presents the $M_{\mathrm{ZFC}}(T)/H$,
$M_{\mathrm{FC}}(T)/H$, and $M_{\mathrm{TRM}}(T)/H$ curves measured
at an applied field of 0.1 Oe. $M_{\mathrm{ZFC}}(T)$ exhibits a
maximum at $T_{f}\approx30$ K in agreement with a previously
reported result \cite{wang}. The inset shows the ZFC and FC curves for 0.1 and 
0.5 Oe; a substantial suppression of $M_{\mathrm{FC}}(T)/H$ is seen, whereas $M_{\mathrm{ZFC}}(T)/H$ is virtually unaffected by the increased field strength. As demonstrated in 
Fig. \ref{fig2}(a), a Curie-Weiss behavior is observed at temperatures 
above $T_{SRF}\approx 70$ K. $M_{\mathrm{TRM}}$ appears nonzero below 
$T_{SRF}$ and both $M_{\mathrm{ZFC}}(T)$ and $M_{\mathrm{FC}}(T)$ deviate from
Curie-Weiss behavior suggesting an establishment of ferromagnetic
correlations related to the double-exchange mechanism. However, our $M(H)$ 
measurements in applied fields
up to 3 T and the corresponding Arrot plots (Fig. \ref{fig2}(b)) show no 
indication of spontaneous magnetization. These results imply that the 
ferromagnetism appearing below $T_{SRF}$ is of short-range order, i.e. 
clusters of short-range FM correlations develop below $T_{SRF}$. Similar 
magnetic properties have previously been reported for
$\mathrm{(Tb_{1/3}La_{2/3})Ca_{1/3}MnO_{3}}$ wherein short-range
ferromagnetic correlations above $T_{f}$ were evidenced from
magnetic and small-angle neutron scattering measurements
\cite{teresa}.\\
It is observed in our $m(t)$ measurements that
$\mathrm{Y_{0.7}Ca_{0.3}MnO_{3}}$ exhibits logarithmically slow dynamics at 
all temperatures below $T_{SRF}$. The system does not
reach equilibrium on time scales up to $10^{4}$ s even at
temperatures far above $T_{f}$. Furthermore, together with the
long-time relaxation, aging effects \cite{lundgren} can also be
seen not only below but also well above $T_{f}$. Fig.
\ref{fig3} displays the relaxation rates $S(t)=1/H(\partial
m(t)/\partial\log t)$ derived from the ZFC $m(t)$ curves recorded at (a)
$T_m=27K <T_{f}$ and (b) $T_m=45K >T_{f}$; $H=0.1$ Oe and 
$t_w$=0s (10s) and 3000s. The figure shows that $S(t)$ attains a characteristic aging 
maximum at an observation time close to $t_{w}$, where there exists an 
inflection point in the corresponding $m(t)$ curves; a similar non-equilibrium 
behavior has been observed in a variety of frustrated and disordered systems 
including some other perovskite compounds\cite{nam,roland,nam2,mira}. In
spin-glasses, the aging effect can be interpreted within the
droplet model \cite{fisher} by associating the maximum in the relaxation
rate to a crossover between a quasi-equilibrium
dynamic regime at short observation times ($t\ll t_{w}$) and
a non-equilibrium regime at long observation times ($t\gg
t_{w}$). Results from Monte Carlo simulations for two- and
three-dimensional Ising spin-glass systems also show that the relaxation 
rate $S(t)$ exhibits a maximum at $t\approx t_{w}$\cite{Andersson}.\\ 
The non-equilibrium dynamics observed at temperatures above $T_{f}$ is 
probably caused by random dipolar interactions between the ferromagnetic 
clusters. In this region, the relaxation time of the system may remain finite 
although it is much larger than the time scales employed in our experiments. 
In fact, in original SG systems, aging effects are found also at temperatures 
above $T_{g}$ at time scales shorter than the maximum relaxation time of
the system \cite{nordblad0}.\\
In both two and three dimensional spin glasses, temperature cycling 
experiments have shown that aging states are virtually unaffected by a 
negative $\Delta T$, while re-initialization occurs for a sufficiently
large positive $\Delta T>0$ \cite{nordblad}. On the other hand, for 
frustrated ferromagnetic systems, re-initialization occurs irrespective of 
the sign of $\Delta T$ \cite{nam2,nordblad,Vincent}. This difference can be 
used to distinguish a spin glass from other frustrated magnets.
$S(t)$ measured at
$T_{m}=27$ K with $\Delta T$=0, -3 and -5 K and $t_{w2}$=0s  are 
indistinguishable from each other, evidencing spin glass behavior below $T_{f}$; the corresponding experiments above $T_{f}$ at $T_{m}=45$ K give the same result. As shown in Fig. \ref{fig4} (a) and (b), if a long wait time $t_{w2}$ is added, a small but noticeable 
reinitialization occurs, in accordance with the behavior of ordinary spin 
glasses\cite{Granberg}. The spin configuration of the aging state at 
$T_m$ seems to be frozen in as the temperature is lowered. In the positive 
cycling experiment one notices an increase of S(t) at short time scales 
indicating re-initialization of the configuration.\\
At $T>T_{f}$ (Fig. \ref{fig4}(b)), the $S(t)$ curves measured at 45 K with 
$\Delta T$=0, -3, -5, and +5 K look very  very similar to the $T_{m}=27$ K 
curves. The material still exhibits a characteristic SG behavior, strikingly  
different from the behavior of a reentrant ferromagnetic 
phase, which could have been anticipated since we observe ferromagnetic ordering 
above $T_f$. However, we deal here with a system that only exhibits short-range
ferromagnetic correlation. In passing, it is worth noting that the magnitude 
of the aging is large below $T_{SRF}$, proving the 
effect to be intrinsic to the material rather than only associated to a 
possible spin disorder at grain boundaries.\\  

\subsection{Ac measurements - memory effects}
\begin{figure}
\centerline{\epsfig{figure=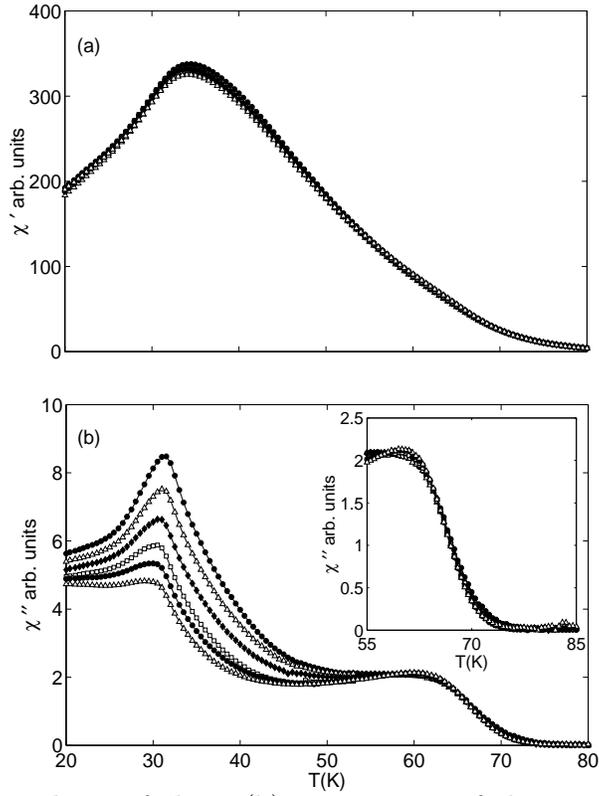,height=10.6cm,width=8cm}}
\caption{In-phase (a) and out-of-phase (b) components of the ac susceptibility for different frequencies: 510, 170, 51, 17, 5.1 and 1.7 Hz; $h_{ac}$=0.01 Oe. The insert shows an enlargement of the 55-85K region.}
\label{fig5}
\end{figure}

\begin{figure}
\centerline{\epsfig{figure=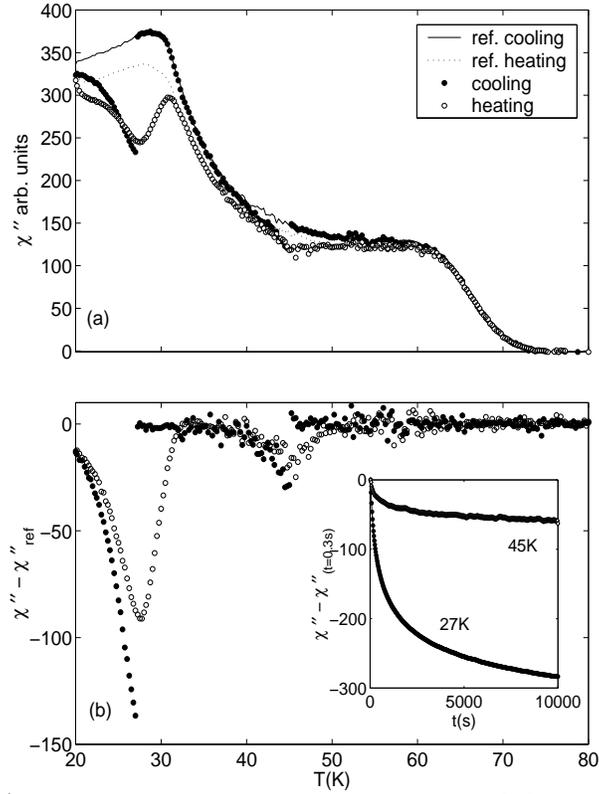,height=10.6cm,width=8cm}}
\caption{(a) $\chi \prime\prime$ and (b) $\chi \prime\prime$ - $\chi \prime\prime_{ref}$ vs. temperatures measured during cooling (filled circles and line) or on re-heating (open circles and dotted line). In the case of the curve with filled circles, the sample was kept 10000s at 45 and 27K during cooling. The insert shows the corresponding relaxation of $\chi \prime\prime$ - $\chi \prime\prime_{(t=0.3s)}$ vs. time.}
\label{fig6}
\end{figure}

\begin{figure}
\centerline{\epsfig{figure=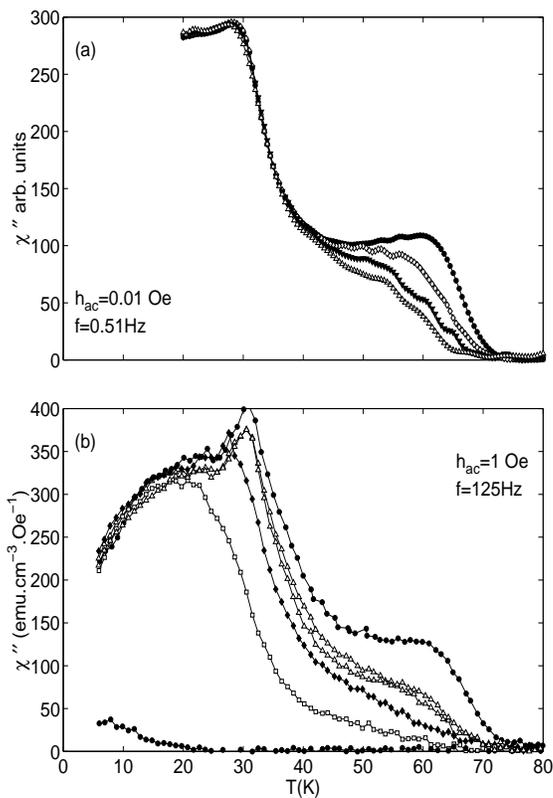,height=10.6cm,width=8cm}}
\caption{Temperature dependence of $\chi \prime\prime$ when superimposing different dc fields. (a) shows the results for small dc fields: 0,1,2 and 5 Oe; (b) for high ones: 0,10,100,300,1000 and 10000 Oe.}
\label{fig7}
\end{figure}

Fig. \ref{fig5} shows the (a) in-phase and (b) out-of-phase components of the 
ac susceptibility vs. temperature for different frequencies.  As seen on 
Fig. \ref{fig5}(b) and insert, the ferromagnetic onset is frequency 
independent. Below $T\approx$60K, the out-of-phase component decreases with 
decreasing frequency. Further decreasing the temperature, there is a frequency 
dependent maximum in the in-phase component and a corresponding but more 
pronounced frequency dependence of the out-of phase component. Using these 
data it is possible to define a freezing temperature $T_f$ that decreases 
with decreasing frequency. Employing the position of maximum slope in the 
out-of-phase component as definition of the freezing temperature at 
observation time $\tau$=1/$\omega$, we have analyzed the data according to 
possible dynamic scaling scenarios. The physically most plausible parameters 
were obtained using activated dynamics and a finite critical temperature, 
$$ln(\tau) \approx - (\frac{1}{T_f})\times[(T_f - T_g)/T_g]^\gamma$$ with $\gamma$=0.87 and $T_g$=28.9K.
However, a microscopic relaxation time of order 1 s was encountered. 
Analyzing the data according to conventional critical slowing down\cite{hohen} 
resulted in rather poor fits. Also, analyses according to Ahrenius or 
generalized Ahrenius slowing down of the dynamics gave no or unphysical 
parameters. There are thus strong indications of the existence of a finite 
phase transition temperature, albeit not necessarily to a conventional low 
temperature spin glass phase.\\
To further characterize the low and high temperature regions, memory effects 
were investigated both below and above $T_f$ using the relaxation of the 
out-of phase component of the ac susceptibility. When cooling from
above $T_{SRF}$ ($T_{g}$ in a conventional SG case), a halt at constant 
temperature $T_{h}<T_{SRF}$ is made during $t_{h}$, allowing the system to 
relax towards its equilibrium state at $T_{h}$; both components of the 
ac susceptibility then decay in magnitude. In spin glasses, this equilibrated 
state becomes frozen in on further lowering the temperature, and is retrieved 
on re-heating the system to $T_{h}$.
The weak low frequency ac-field employed in this kind of experiments does not
affect the non-equilibrium processes intrinsic to the sample, but only
works as a non-perturbing probe of the system. A memory effect is here clearly 
observed, Fig. \ref{fig6}, not only at $T$=27K but, surprisingly, also at 
45K. The memory dip appears even more clearly  when substrating the references
curves as in Fig. \ref{fig6}(b). The insert shows the out-of-phase component 
of the ac susceptibility recorded vs time at T=27 and 45 K after direct cooling
from above $T_{SRF}$. As already seen in the memory plot, the relaxation is 
comparably smaller at $T=45K$. One notices that the measured relaxation at 
both temperatures is larger than in the memory experiment; this is because 
the ac susceptibility in this case was recorded directly after cooling the 
system from the reference temperature above $T_{SRF}$ to the measurement temperature.\\
A superimposed dc field affects the ac susceptibility of spin glasses in a 
peculiar and signifying way. The in-phase component is significantly suppressed,
but only at temperatures above the freezing temperature $T_f(H,\omega)$; and 
the onset of the out-of-phase component is suppressed to lower temperatures, 
but also remains unaffected at lower temperatures\cite{Mattsson}! In Fig. 
\ref{fig7} we have plotted $\chi''$ in different superposed dc-fields. Fig. 
\ref{fig7}(a), shows that the near-$T_{SRF}$ magnitude and onset is  
fragile with respect to even a weak superimposed dc-field. $\chi''$ is 
substantially affected by a dc-field of only 1 Oe, and is suppressed to lower 
temperatures already at 5 Oe. At lower temperatures around $T_f$ the 
out-of-phase component remains unaffected by the dc-field. At larger dc-fields,
Fig. \ref{fig7} (b) shows that also the freezing temperature becomes 
suppressed, but that the ac susceptibility below $T_f$ remains unaffected in 
a spin glass like fasion\cite{Mattsson}.  

\section{CONCLUSION}
The magnetic response of $\mathrm{Y_{0.7}Ca_{0.3}MnO_{3}}$ becomes governed 
by short range ferromagnetic correlations at temperatures below 
$T_{SRF}\approx70$ K. Above $T_{f}$, a surprisingly SG-like state is 
observed, featuring aging and memory effects. The non-equilibrium dynamics 
above $T_{f}$ may be attributed to a thermally activated redistribution of 
ferromagnetically ordered clusters and the random  dipolar interaction 
amongst their magnetic moments. This state seemingly evolves into a 
conventional spin-glass state at temperatures below $T_{f}$.

\acknowledgments

Financial support from the Swedish Natural Science Research
Council (NFR) is acknowledged. This work is also sponsored by
Sida/SAREC - and the ISP of Uppsala University and partially by a
Vietnam's Grant on Basic Research. The authors thank Dr. P. V.
Phuc and N. D. Van for the x-ray measurements and phase analyses. Special thanks
are due to Prof. N. Chau, Dr. N. L. Minh, and Dr. B. T. Cong for
their help in sample preparation. The authors are also extremely grateful to Dr. P. J\"onsson for her help.

\end{document}